\documentclass[10pt,conference]{IEEEtran}

\usepackage{cite,graphicx,psfrag,amsmath,amssymb,url}

\def\mid{{\mathop{\rm mid}}}
\def\hi{{\mathop{\rm hi}}}
\def\lo{{\mathop{\rm lo}}}
\def\ub{{\mathop{\rm ub}}}

\def\os{{\mathop{\rm CC}}}
\def\argmin{{\mathop{\arg\,\min}}}
\def\order{O}

\def\s{\mbox{'s}}
\def\for{\mbox{ for }}
\def\st{\mbox{ such that }}
\def\total{{\mathop{\rm tot}}}
\def\power{{\mathop{\rm pow}}}
\def\rhot{{\rho_{\total}}}
\def\rhop{{\rho_{\power}}}

\def\AlphaSet{{A}}
\def\BetaSet{{B}}
\def\GammaSet{{\mathit{\Gamma}}}
\def\DeltaSet{{\mathit{\Delta}}}
\def\algd{D}
\def\nodeset{\eta}

\def\definedas{\triangleq}
\def\das{\triangleq}
\def\smalll{{\mbox{\scriptsize \boldmath $l$}}}
\def\boldl{{\mbox{\boldmath $l$}}}
\def\letterl{l}
\def\posinf{{\infty}}

\def\p{{\mbox{\boldmath $p$}}}

\def\I{{\mathcal I}}
\def\letterk{{\kappa}}
\def\P{{\mathcal P}}
\def\PS{{\mathcal P}^*}
\def\R{{\mathbb R}}
\def\Rp{{\mathbb R}_+}
\def\S{{\mathcal S}}
\def\X{{\mathcal X}}
\def\Z{{\mathbb Z}}

\def\boldlvarphi{{\boldl^{(\varphi)}}}
\def\boldlchi{{\boldl^{(\chi)}}}
\def\boldlupN{{\boldl^{N}}}
\newcommand{\defn}[0]{\textit}
\newtheorem{theorem}{Theorem}
\newtheorem{lemma}{Lemma}
\newtheorem{definition}{Definition}

\begin{document}
\title{Twenty (or so) Questions: $D$-ary Length-Bounded Prefix Coding}

\author{Michael~B.~Baer \\
Electronics for Imaging\\
303 Velocity Way\\
Foster City, California  94404\\
Email: Michael.Baer@efi.com}

\maketitle

\begin{abstract}
Efficient optimal prefix coding has long been accomplished via the Huffman algorithm.  However, there is still room for improvement and exploration regarding variants of the Huffman problem.  Length-limited Huffman coding, useful for many practical applications, is one such variant, for which codes are restricted to the set of codes in which none of the $n$ codewords is longer than a given length, $l_{\max}$. Binary length-limited coding can be done in $O(n l_{\max})$ time and $O(n)$ space via the widely used Package-Merge algorithm and with even smaller asymptotic complexity using a lesser-known algorithm.  In this paper these algorithms are generalized without increasing complexity in order to introduce a minimum codeword length constraint $l_{\min}$, to allow for objective functions other than the minimization of expected codeword length, and to be applicable to both binary and nonbinary codes; nonbinary codes were previously addressed using a slower dynamic programming approach.  These extensions have various applications --- including fast decompression and a modified version of the game ``Twenty Questions'' --- and can be used to solve the problem of finding an optimal code with limited fringe, that is, finding the best code among codes with a maximum difference between the longest and shortest codewords.  The previously proposed method for solving this problem was nonpolynomial time, whereas solving this using the novel linear-space algorithm requires only $O(n (l_{\max}- l_{\min})^2)$ time, or even less if $l_{\max}- l_{\min}$ is not $O(\log n)$.
\end{abstract}

\section{Introduction} 
\label{intro} 
The parlor game best known as ``Twenty Questions'' has a long history
and a broad appeal.  It was used to advance the plot of Charles
Dickens' \textit{A Christmas Carol}\cite{Dic}, in which it is called
``Yes and No,'' and it was used to explain information theory in
Alfr\'{e}d R\'{e}nyi's \textit{A Diary on Information Theory}\cite{Reny},
in which it is called ``Bar-kochba.''  The two-person game begins with an
answerer thinking up an object and then being asked a series of
questions about the object by a questioner.  These questions must be
answered either ``yes'' or ``no.''  Usually the questioner can ask at
most twenty questions, and the winner is determined by whether or not
the questioner can sufficiently surmise the object from these questions.

Many variants of the game exist --- both in name and in rules.  A
recent popular variant replaces the questioner with an electronic
device\cite{Cass}.  The answerer can answer the device's questions
with one of four answers --- ``yes,'' ``no,'' ``sometimes,'' and
``unknown.''  The game also differs from the traditional game in that
the device often needs to ask more than twenty questions.  If the
device needs to ask more than the customary twenty questions, the
answerer can view this as a partial victory, since the device has not
answered correctly given the initial twenty.  However, the device 
eventually gives up after $25$ questions if it cannot guess the
questioner's object.

Consider a short example of such a series of questions, with only
``yes,'' ``no,'' and ``sometimes'' as possible answers.  The object to
guess is one of the seven Newtonian colors\cite{Newt}, which we choose
to enumerate as follows:
\begin{enumerate}
\item Green (G)
\item Yellow (Y)
\item Red (R)
\item Orange (O)
\item Indigo (I)
\item Violet (V)
\item Blue (B).
\end{enumerate}
A first question we ask might be, ``Is the color seen as a warm
color?''  If the answer is ``sometimes,'' the color is green.  If it
is ``yes,'' it is one of colors $2$ through~$4$.  If so, we then ask,
``Is the color considered primary?''  ``Sometimes'' implies yellow,
``yes'' implies red, and ``no'' implies orange.  If the color is not
warm, it is one of colors $5$ through $7$, and we ask whether the
color is considered purple, a different question than the one for
colors $2$ through $4$.  ``Sometimes'' implies indigo, ``yes'' implies
violet, and ``no'' implies blue.  Thus we can distinguish the seven
colors with an average of $2-p_1$ questions if $p_1$ is the
probability that color in question is green.

\begin{figure*}[ht]
\centering
\psfrag{0}{\small $0_3$}
\psfrag{10}{\small $10_3$}
\psfrag{11}{\small $11_3$}
\psfrag{12}{\small $12_3$}
\psfrag{20}{\small $20_3$}
\psfrag{21}{\small $21_3$}
\psfrag{22}{\small $22_3$}
\psfrag{sin 1}{\tiny 1) G}
\psfrag{sin 2}{\tiny 2) Y}
\psfrag{sin 3}{\tiny 3) R}
\psfrag{sin 4}{\tiny 4) O}
\psfrag{sin 5}{\tiny 5) I}
\psfrag{sin 6}{\tiny 6) V}
\psfrag{sin 7}{\tiny 7) B}
\psfrag{atop}{\small $\alpha_{l_{\max}}=3$}
\psfrag{amid}{\small $\alpha_{l_{\max}-1}=2$}
\psfrag{abot}{\small $\alpha_{l_{\max}-2}=0$}
\includegraphics{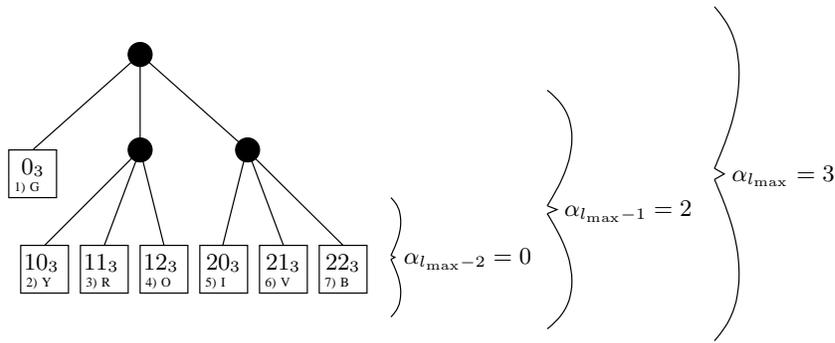}
\caption{A monotonic code tree for $n=7$ and $\algd=3$ with $\boldl =
(1, 2, 2, 2, 2, 2, 2)$: Each leaf contains the trinary output code,
the corresponding object number, and the initial for the corresponding
color as in Section~\ref{intro}.  The $\alpha_i\s$ are as defined in
Section~\ref{linpen}.}
\label{codetree}
\end{figure*}

This series of questions is expressible using code tree notation,
e.g., \cite{Schw}, in which a tree is formed with each child split
from its parent according to the corresponding output symbol, i.e.,
the answer of the corresponding question.  A code tree corresponding
to the above series of questions is shown in Fig.~\ref{codetree}, where
a left branch means ``sometimes,'' a middle branch means ``yes,'' and
a right branch means ``no.''  The number of answers possible is
referred to by the constant $\algd$ and the tree is a $\algd$-ary
tree.  In this case, $\algd=3$ and the code tree is ternary.  The
number of outputs, $n=7$, is the number of colors.

The analogous problem in prefix coding is as follows: A source (the
answerer) emits input symbols (objects) drawn from the alphabet $\X =
\{ 1, 2, \ldots, n \}$, where $n$ is an integer.  Symbol $i$ has
probability $p_i$, thus defining probability vector $\p=(p_1, p_2,
\ldots, p_n)$.  Only possible symbols are considered for coding and
these are sorted in decreasing order of probability; thus $p_i > 0$
and $p_i \leq p_j$ for every $i>j$ such that $i, j \in \X$.  (Since
sorting is only $\order(n \log n)$ time and $\order(n)$ space, this can
be assumed without loss of generality.)  Each input symbol is encoded
into a codeword composed of output symbols of the $\algd$-ary alphabet
$\{0, 1, \ldots, \algd - 1\}$.  (In the example of colors, $0$
represents ``sometimes,'' $1$ ``yes,'' and $2$ ``no.'')  The codeword
$c_i$ corresponding to input symbol $i$ has length $\letterl_i$, thus
defining length vector $\boldl = (l_1, l_2, \ldots, l_n)$.  In
Fig.~\ref{codetree}, for example, $c_7$ is $22_3$ --- the codeword
corresponding to blue --- so length $l_7 = 2$.  The overall code
should be a prefix code, that is, no codeword $c_i$ should begin with
the entirety of another codeword~$c_j$.  In the game, equivalently, we
should know when to end the questioning, this being the point at which
we know the answer.

For the variant introduced here, all codewords must have lengths lying
in a given interval [$l_{\min}$,$l_{\max}$].  In the example of the
device mentioned above, $l_{\min} = 20$ and $l_{\max} = 25$.  A more
practical variant is the problem of designing a data codec which is
efficient in terms of both compression ratio and coding speed.  Moffat
and Turpin proposed a variety of efficient implementations of prefix
encoding and decoding in \cite{MoTu97}, each involving table lookups
rather than code trees.  They noted that the length of the longest
codeword should be limited for computational efficiency's sake.
Computational efficiency is also improved by restricting the overall
range of codeword lengths, reducing the size of the tables and the
expected time of searches required for decoding.  Thus, one might wish
to have a minimum codeword size of, say, $l_{\min}=16$ bits and a
maximum codeword size of $l_{\max}=32$ bits ($\algd=2$).  If expected
codeword length for an optimal code found under these restrictions is
too long, $l_{\min}$ can be reduced and the algorithm rerun until the
proper trade-off between coding speed and compression ratio is found.

A similar problem is one of determining opcodes of a
microprocessor designed to use variable-length opcodes, each a certain
number of bytes ($\algd=256$) with a lower limit and an upper limit to
size, e.g., a restriction to opcodes being 16, 24, or 32 bits long
($l_{\min}=2$, $l_{\max}=4$).  This problem clearly falls within the
context considered here, as does the problem of assigning video
recorder scheduling codes; these human-readable decimal codes
($\algd=10$) have lower and upper bounds on their size, such as
$l_{\min} = 3$ and $l_{\max} = 8$, respectively.  

Other problems of interest have $l_{\min} = 0$ and are thus length
limited but have no practical lower bound on length\cite[p.~396]{WMB}.
Yet other problems have not fixed bounds but a constraint on the
difference between minimum and maximum codeword length, a quantity
referred to as fringe \cite[p.~121]{Abr01}.  As previously noted,
large fringe has a negative effect of the speed of a decoder.  In
Section~\ref{conclusion} of this paper we discuss how to find such
codes.

Note that a problem of size $n$ is trivial for certain values of
$l_{\min}$ and~$l_{\max}$.  If $l_{\min} \geq \log_\algd n$, then all
codewords can have $l_{\min}$ output symbols, which, by any reasonable
objective, forms an optimal code.  If
$l_{\max} < \log_\algd n$, then we cannot code all input symbols and the
problem, as presented here, has no solution.  Since only other values
are interesting, we can assume that $n \in
(\algd^{l_{\min}},\algd^{l_{\max}}]$.  For example, for the modified
form of Twenty Questions, $\algd=4$, $l_{\min} = 20$, and $l_{\max} =
25$, so we are only interested in problems where $n \in
(2^{40},2^{50}]$.  Since most instances of Twenty Questions have fewer
possible outcomes, this is usually not an interesting problem after
all, as instructive as it is.  In fact, the fallibility of the
answerer and ambiguity of the questioner mean that a decision tree
model is not, strictly speaking, correct.  For example, the
aforementioned answers to questions about the seven colors are
debatable.  The other applications of \defn{length-bounded} prefix coding
mentioned previously, however, do fall within this model.

If we either do not require a minimum or do not require a maximum, it
is easy to find values of $l_{\min}$ or $l_{\max}$ which do not limit
the problem.  As mentioned, setting $l_{\min} = 0$ results in a
trivial minimum, as does $l_{\min} = 1$.  Similarly, setting $l_{\max}
= n$ or using the hard upper bound $l_{\max} = \lceil (n-1)/(\algd-1)
\rceil$ results in a trivial maximum value.  In the case of trivial
maximum values, one can actually minimize expected codeword length in
linear time given sorted inputs.  This is possible because, at each
stage in the standard Huffman coding algorithm, the set of Huffman
trees is an optimal \defn{forest} (set of trees)\cite{HuTu}.  We
describe the linear-time algorithm in Section~\ref{refine}.  

If both minimum and maximum values are trivial, 
Huffman coding~\cite{Huff} yields a prefix code minimizing expected codeword
length
\begin{equation}
\sum_{i=1}^n p_i \letterl_i .
\label{huff}
\end{equation}
The conditions necessary and sufficient for the existence of a 
prefix code with length vector $\boldl$ are
the integer constraint, $\letterl_i \in \Z_+$, and the
Kraft (McMillan) inequality~\cite{McMi},
\begin{equation}
\letterk(\boldl) \definedas \sum_{i=1}^n \algd^{-\letterl_i}\leq 1 .
\label{kraft}
\end{equation}
Finding values for $\boldl$ is
sufficient to find a corresponding code, as a code tree with the
optimal length vector can be built from sorted codeword lengths in
$\order(n)$ time and space. 

It is not always obvious that we should minimize the expected
number of questions $\sum_i p_i l_i$ (or, equivalently, the expected
number of questions in excess of the first~$l_{\min}$,
\begin{equation}
\sum_{i=1}^n p_i (l_i-l_{\min})^+
\end{equation}
where $x^+$ is $x$ if $x$ is positive, $0$ otherwise).  Consider the
example of video recorder scheduling codes.  In such an application,
one might instead want to minimize mean square distance
from~$l_{\min}$,
$$\sum_{i=1}^n p_i (l_i-l_{\min})^2.$$  
We generalize and investigate how to
minimize the value
\begin{equation}
\sum_{i=1}^n p_i \varphi(l_i-l_{\min})
\label{penalty}
\end{equation}
under the above constraints
for any \defn{penalty function} $\varphi(\cdot)$ convex and increasing
on~$\Rp$.  Such an additive measurement of cost is called a
\defn{quasiarithmetic penalty}, in this case a convex quasiarithmetic
penalty.

One such function $\varphi$ is $\varphi(\delta) = (\delta+l_{\min})^2$, a
quadratic value useful in optimizing a communications delay
problem~\cite{Larm}.  Another function, $\varphi(\delta) =
\algd^{t(\delta+l_{\min})}$ for $t>0$, can be used to minimize the
probability of buffer overflow in a queueing system\cite{Humb2}.

Mathematically stating the length-bounded problem,
$$
\begin{array}{ll}
\mbox{Given } & \p = (p_1, \ldots, p_n),~p_i > 0; \\
& \algd \in \{2, 3, \ldots\}; \\
& \mbox{convex, monotonically increasing } \\
& \varphi: \Rp \rightarrow \Rp \\
\mbox{Minimize} \,_{\{\smalll\}} &
\sum_i p_i \varphi(l_i-l_{\min}) \\
\mbox{subject to } & \sum_i \algd^{-l_i} \leq 1; \\
& l_i \in \{l_{\min},l_{\min}+1,\ldots,l_{\max}\}.
\end{array}
$$ Note that we need not assume that probabilities $p_i$ sum to $1$;
they could instead be arbitrary positive weights.  

Thus, in this paper, given a finite $n$-symbol input alphabet with an
associated probability vector $\p$, a $\algd$-symbol output alphabet
with codewords of lengths $[l_{\min},l_{\max}]$ allowed, and a
constant-time-calculable convex penalty function $\varphi$, we describe an
$\order(n(l_{\max}-l_{\min}))$-time $\order(n)$-space algorithm for
constructing a $\varphi$-optimal code, and sketch an even less complex
reduction for the most convex penalty function,
$\varphi(\delta)=\delta$, minimization of expected codeword length.
In the next section, we present a brief review of the relevant
literature.  In Section~\ref{nodeset}, we extend to $\algd$-ary codes
an alternative to code tree notation first presented in \cite{Larm}.
This notation aids in solving the problem in question by reformulating
it as an instance of the $\algd$-ary Coin Collector's problem,
presented in Section~\ref{cc} as an extension of the (binary) Coin
Collector's problem\cite{LaHi}.  An extension of the Package-Merge
algorithm solves this problem; we introduce the reduction and
resulting algorithm in Section~\ref{algorithm}.  We make it
$\order(n)$ space in Section~\ref{linear} and refine it in
Section~\ref{refine}.  The alternative approach for the expected
length problem of minimizing (\ref{huff}) --- i.e.,
$\varphi(\delta)=\delta$ --- is often faster; this approach is
sketched in Section~\ref{linpen}.  Algorithmic modifications,
applications, possible extensions of this work are discussed in
Section~\ref{conclusion}.

\section{Prior Work}
\label{prior}

Reviewing how the problem in question differs from binary Huffman
coding:
\begin{enumerate}
\item It can be nonbinary, a case considered by Huffman in his original
paper\cite{Huff};
\item There is a maximum codeword length, a restriction previously
considered, e.g., \cite{Itai} in $\order(n^3 l_{\max} \log \algd)$
time \cite{GoWo} and $\order(n^2 \log \algd)$ space, but solved
efficiently only for binary coding, e.g., \cite{LaHi} in $\order(n
l_{\max})$ time $\order(n)$ space and most efficiently in \cite{Schi};
\item There is a minimum codeword length, a novel restriction;
\item The penalty can be nonlinear, a modification previously
considered, but only for binary coding, e.g., \cite{Baer06}.
\end{enumerate}
There are several methods for finding optimal codes for various
constraints and various types of optimality; we review the three most
common families of methods here.  Note that other methods fall outside
of these families, such as a linear-time method for finding minimum
expected length codewords for a uniform distribution with a given
fringe\cite{DePe}.  (This differs from the limited-fringe problem of
Section~\ref{conclusion}, in which the distribution need not be
uniform and fringe is upper-bounded, not fixed.)

The first and computationally simplest of these are Huffman-like
methods, originating with Huffman in 1952\cite{Huff} and discussed in,
e.g., \cite{ChTh}.  Such algorithms are generally linear time given
sorted weights and thus $\order(n \log n)$ time in general.  These are
useful for a variety of problems involving penalties in linear,
exponential, or minimax form, but not for other nonlinearities nor for
length-limited coding.  More complex variants of this
algorithm are used to find optimal alphabetic codes, that is, codes
with codewords constrained to be in a given lexicographical order.
These variants are in the Hu-Tucker family of
algorithms\cite{HuTu,GaWa,HKT}, which, at $\order(n \log n)$ time and
$\order(n)$ space\cite{Knu31}, are the most efficient algorithms known
for solving such problems (although some instances can be solved in
linear time\cite{HuMo,HLM}).

The second type of method, dynamic programming, is also conceptually
simple but much more computationally complex.  Gilbert and Moore
proposed a dynamic programming approach in 1959 for finding optimal
alphabetic codes, and, unlike the Hu-Tucker algorithm, this approach
is readily extensible to search trees\cite{Knu71}.  Such an approach
can also solve the nonalphabetic problem as a special case, e.g.,
\cite{HuTa,Gare,Itai}, since any probability vector satisfying $p_i
\leq p_j$ for every $i>j$ has an optimal alphabetic code that
optimizes the nonalphabetic case.  A different dynamic programming
approach can be used to find optimal ``1''-ended codes\cite{ChGo} and
optimal codes with unequal letters costs\cite{GoRo}.  Itai\cite{Itai}
used dynamic programming to optimize a large variety of coding and
search tree problems, including nonbinary length-limited 
coding, which is done with $\order(n^2 l_{\max} \log \algd)$ time and
$\order(n^2 \log \algd)$ space by a reduction to the alphabetic case.
We reduce complexity significantly in this paper.

The third family is that of Package-Merge-based algorithms, and this
is the type of approach we use for the generalized algorithm
considered here.  Introduced in 1990 by Larmore and
Hirschberg\cite{LaHi}, this approach is most often used for binary
length-limited linear-penalty Huffman coding, although it has been
extended for application to binary alphabetic codes \cite{LaPr2} and
to binary convex quasiarithmetic penalty functions \cite{Baer06}.  The
algorithms in this approach generally have $\order(n l_{\max})$-time
$\order(n)$-space complexity, although space complexity can vary by
application and implementation, and the alphabetic variant and some
nonquasiarithmetic (and thus nonlinear) variants have slightly higher
time complexity ($\order(n l_{\max} \log n)$).

To use this approach for nonbinary coding with a lower bound on
codeword length, we need to alter the approach, generalizing to the
problem of interest.  The minimum size constraint on codeword length
requires a relatively simple change of solution range.  The nonbinary
coding generalization is a bit more involved; it requires first
modifying the Package-Merge algorithm to allow for an arbitrary
numerical base (binary, ternary, etc.), then modifying the coding
problem to allow for a provable reduction to the modified
Package-Merge algorithm.  At times ``dummy'' inputs are added in order
to assist in finding an optimal nonbinary code.  In order to make the
algorithm precise, the $\order(n(l_{\max}-l_{\min}))$-time
$\order(n)$-space algorithm, unlike some other
implementations\cite{LaHi}, minimizes \defn{height} (that is, maximum
codeword length) among optimal codes (if multiple optimal codes
exist).

\section{Nodeset Notation}
\label{nodeset}

Before presenting an algorithm for optimizing the above problem, we
introduce a notation for codes that generalizes one first presented
in~\cite{Larm} and modified in \cite{Baer06}.  Nodeset notation, an
alternative to code tree notation, has previously been used for binary
alphabets, but not for general $\algd$-ary alphabet coding, thus the
need for generalization.

\textit{The key idea:} Each node $(i,l)$ represents both the share of
the penalty (\ref{penalty}) (\defn{weight}) and the (scaled) share of
the Kraft sum (\ref{kraft}) (\defn{width}) assumed
for the $l$th bit of the $i$th codeword.  By showing that total weight
is an increasing function of the penalty and that there is a
one-to-one correspondence between an optimal code and a corresponding optimal
nodeset, we reduce the problem to an
efficiently solvable problem, the Coin Collector's problem.

In order to do this, we first need to make a modification to the
problem analogous to one Huffman made in his original nonbinary
solution.  We must in some cases add a ``dummy'' input or ``dummy''
inputs of infinitesimal probability $p_i = \epsilon > 0$ to the
probability vector to assure that the optimal code has the Kraft
inequality satisfied with equality, an assumption underlying both the
Huffman algorithm and ours.  The positive probabilities of these dummy
inputs mean that codes obtained could be slightly suboptimal, but we
later specify an algorithm where $\epsilon = 0$, obviating this
concern.  

As with traditional Huffman coding\cite{Huff}, the number of dummy
values needed is $(\algd-n) \bmod{(\algd-1)}$, where $$x\bmod{y} \das
x - y \lfloor x/y \rfloor$$ for all integers $x$ (not just nonnegative
integers).  Such dummy inputs allow us to assume that the optimal tree
(for real plus dummy items) is an optimal full tree (i.e., that
$\letterk(\boldl)=1$, where $\letterk$ is as defined in
(\ref{kraft})).  For sufficiently small $\epsilon$, the code will be
identical to that for $\epsilon=0$, and, as in traditional Huffman
coding, nondummy codewords are identical to the codewords of an
optimal code for the original input distribution.  We can thus assume for
our algorithm that $\letterk(\boldl)=1$ and $n \bmod{(\algd-1)} \equiv
1$.

\begin{figure*}[ht]
\centering
\psfrag{l (level)}{$l$ (level)}
\psfrag{i (item)}{$i$ (input symbol)}
\psfrag{(width)}{$\rho$ (width)}
\psfrag{1m1}{\scriptsize $\mu(1,2) = p_1$}
\psfrag{2m1}{\scriptsize $\mu(2,2) = p_2$}
\psfrag{3m1}{\scriptsize $\mu(3,2) = p_3$}
\psfrag{4m1}{\scriptsize $\mu(4,2) = p_4$}
\psfrag{5m1}{\scriptsize $\mu(5,2) = p_5$}
\psfrag{6m1}{\scriptsize $\mu(6,2) = p_6$}
\psfrag{7m1}{\scriptsize $\mu(7,2) = p_7$}
\psfrag{1m2}{\scriptsize $\mu(1,3) =3p_1$}
\psfrag{2m2}{\scriptsize $\mu(2,3) =3p_2$}
\psfrag{3m2}{\scriptsize $\mu(3,3) =3p_3$}
\psfrag{4m2}{\scriptsize $\mu(4,3) =3p_4$}
\psfrag{5m2}{\scriptsize $\mu(5,3) =3p_5$}
\psfrag{6m2}{\scriptsize $\mu(6,3) =3p_6$}
\psfrag{7m2}{\scriptsize $\mu(7,3) =3p_7$}
\psfrag{1m3}{\scriptsize $\mu(1,4) =5p_1$}
\psfrag{2m3}{\scriptsize $\mu(2,4) =5p_2$}
\psfrag{3m3}{\scriptsize $\mu(3,4) =5p_3$}
\psfrag{4m3}{\scriptsize $\mu(4,4) =5p_4$}
\psfrag{5m3}{\scriptsize $\mu(5,4) =5p_5$}
\psfrag{6m3}{\scriptsize $\mu(6,4) =5p_6$}
\psfrag{7m3}{\scriptsize $\mu(7,4) =5p_7$}
\psfrag{1r1}{\scriptsize $\rho(1,2) = \frac{1}{9}$}
\psfrag{1r2}{\scriptsize $\rho(2,2) = \frac{1}{9}$}
\psfrag{1r3}{\scriptsize $\rho(3,2) = \frac{1}{9}$}
\psfrag{1r4}{\scriptsize $\rho(4,2) = \frac{1}{9}$}
\psfrag{1r5}{\scriptsize $\rho(5,2) = \frac{1}{9}$}
\psfrag{1r6}{\scriptsize $\rho(6,2) = \frac{1}{9}$}
\psfrag{1r7}{\scriptsize $\rho(7,2) = \frac{1}{9}$}
\psfrag{2r1}{\scriptsize $\rho(1,3) = \frac{1}{27}$}
\psfrag{2r2}{\scriptsize $\rho(2,3) = \frac{1}{27}$}
\psfrag{2r3}{\scriptsize $\rho(3,3) = \frac{1}{27}$}
\psfrag{2r4}{\scriptsize $\rho(4,3) = \frac{1}{27}$}
\psfrag{2r5}{\scriptsize $\rho(5,3) = \frac{1}{27}$}
\psfrag{2r6}{\scriptsize $\rho(6,3) = \frac{1}{27}$}
\psfrag{2r7}{\scriptsize $\rho(7,3) = \frac{1}{27}$}
\psfrag{3r1}{\scriptsize $\rho(1,4) = \frac{1}{81}$}
\psfrag{3r2}{\scriptsize $\rho(2,4) = \frac{1}{81}$}
\psfrag{3r3}{\scriptsize $\rho(3,4) = \frac{1}{81}$}
\psfrag{3r4}{\scriptsize $\rho(4,4) = \frac{1}{81}$}
\psfrag{3r5}{\scriptsize $\rho(5,4) = \frac{1}{81}$}
\psfrag{3r6}{\scriptsize $\rho(6,4) = \frac{1}{81}$}
\psfrag{3r7}{\scriptsize $\rho(7,4) = \frac{1}{81}$}
\psfrag{1}{$1$}
\psfrag{2}{$2$}
\psfrag{3}{$3$}
\psfrag{4}{$4$}
\psfrag{5}{$5$}
\psfrag{6}{$6$}
\psfrag{7}{$7$}
\resizebox{13cm}{!}{\includegraphics{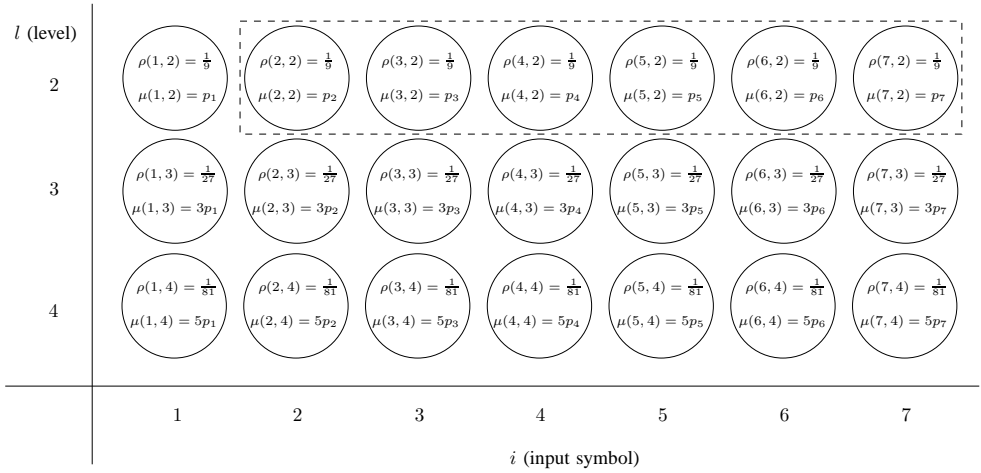}}
\caption{The set of nodes $I$ with widths $\{\rho(i,l)\}$ and weights
$\{\mu(i,l)\}$ for $\varphi(\delta) = \delta^2$, $n=7$, $\algd=3$,
$l_{\min} = 1$, $l_{\max}=4$}
\label{nodesetnum}
\end{figure*}

With this we now present \defn{nodeset} notation:

\begin{definition} A \defn{node} is an ordered pair of integers $(i, l)$ such
  that $i \in \{1,\ldots,n\}$ and $l \in
  \{l_{\min}+1,\ldots,l_{\max}\}$.  Call the set of all possible
  nodes~$I$.  This set can be arranged in an $n \times
  (l_{\max}-l_{\min})$ grid, e.g., Fig.~\ref{nodesetnum}.  The set of
  nodes, or \defn{nodeset}, corresponding to input symbol $i$ (assigned
  codeword~$c_i$ with length~$l_i$) is the set of the
  first~$l_i-l_{\min}$ nodes of column~$i$, that is,
  $\nodeset_\smalll(i) \definedas \{(j,l)~|~j=i,~l \in
  \{l_{\min}+1,\ldots,l_i\}\}$.  The nodeset corresponding to length
  vector~$\boldl$ is $\nodeset(\boldl) \definedas \bigcup_i
  \nodeset_\smalll(i)$; this corresponds to a set of $n$ codewords, a
  code.  Thus, in Fig.~\ref{nodesetnum}, the dashed line surrounds a
  nodeset corresponding to $\boldl = (1, 2, 2, 2, 2, 2, 2)$.  We say a
  node $(i,l)$ has \defn{width} $\rho(i,l) \definedas \algd^{-l}$ and
  \defn{weight} $\mu(i,l) \definedas p_i \varphi(l-l_{\min}) - p_i
  \varphi(l-l_{\min}-1)$, as shown in the example in Fig.~\ref{nodesetnum}.
  Note that if $\varphi(l)=l$, $\mu(i,l)=p_i$.
\end{definition}

We must emphasize that the above ``nodes'' are unlike nodes in a
graph; similar structures are sometimes instead called
\defn{tiles}\cite{LaPr1}, but we retain the original, more prevalent term
``nodes.''  Given valid nodeset $N
\subseteq I$, it is straightforward to find the corresponding length
vector and, if it satisfies the Kraft inequality, a code.

\section{The $\algd$-ary Coin Collector's Problem and the Package-Merge Algorithm}
\label{cc}
We find optimal codes by first solving a related problem, the Coin
Collector's problem.  Let $\algd^\Z$ denote the set of all integer
powers of a fixed integer $\algd>1$.  The Coin Collector's problem of
size $m$ considers ``coins'' indexed by $i \in \{1, 2, \ldots, m\}$.
Each coin has a width, $\rho_i \in \algd^\Z$; one can think of width
as coin face value, e.g., $\rho_i = 0.25 = 2^{-2}$ for a quarter
dollar (25 cents).  Each coin also has a weight, $\mu_i \in \R$.  The
final problem parameter is total width, denoted~$\rhot$.  The problem
is then:
\begin{equation}
\begin{array}{ll}
\mbox{Minimize} \,_{\{B \subseteq \{1,\ldots,m\}\}} & \sum_{i \in B} \mu_i  \\
\mbox{subject to } & \sum_{i \in B} \rho_i = \rhot \\
\mbox{where } & m \in \Z_+ \\ 
& \mu_i \in \R \\
& \rho_i \in \algd^\Z \\
& \rhot \in \R_+ .
\end{array} \label{knap}
\end{equation}
We thus wish to choose coins with total width $\rhot$ such that their
total weight is as small as possible.  This problem is an
input-restricted variant of the knapsack problem.  However, given sorted
inputs, a linear-time solution to (\ref{knap}) for $\algd=2$ was
proposed in \cite{LaHi}.  The algorithm in question is called the
\defn{Package-Merge algorithm} and we extend it here to arbitrary~$\algd$.

In our notation, we use $i \in \{1,\ldots,m\}$ to denote both the
index of a coin and the coin itself, and $\I$ to represent the $m$
items along with their weights and widths.  The optimal solution, a
function of total width $\rhot$ and items $\I$, is denoted $\os(\I,\rhot)$
(the optimal coin collection for $\I$ and $\rhot$).  Note
that, due to ties, this need not be unique, but we assume that one of
the optimal solutions is chosen; at the end of Section~\ref{linear},
we discuss how to break ties.

Because we only consider cases in which a solution exists, $\rhot =
\omega \rhop$ for some $\rhop \in \algd^\Z$ and $\omega \in \Z_+$.
Here, assuming $\rhot>0$, $\rhop$ and $\omega$ are the unique pair of
a power of $\algd$ and an integer that is not a multiple of $\algd$,
respectively, which, multiplied, form~$\rhot$.  If $\rhot = 0$,
$\omega$ and $\rhop$ are not used.  Note that $\rhop$ need not be an
integer.

\noindent \textbf{Algorithm variables} \\
At any point in the algorithm, given nontrivial $\I$ and $\rhot$, we use the following definitions: 

\begin{tabular}{rcl}
Remainder & & \\
$\rhop$ & $\definedas$ & the unique $x \in \algd^\Z$ \\
& & such that $\frac{\rhot}{x} \in \Z \backslash \algd\Z$ \\[2pt]
Minimum width & & \\
$\rho^*$ & $\definedas$ & $\min_{i \in \I} \rho_i$ \\
& & (note $\rho^* \in \algd^\Z$) \\[2pt]
Small width set & & \\
$\I^*$ & $\definedas$ & $\{i~|~\rho_i = \rho^*\}$ \\
& & (note $\I^* \neq \emptyset$) \\[2pt]
``First'' item & & \\
$i^*$ & $\definedas$ & $\argmin_{i \in \I^*} \mu_i$ \\[2pt]
``First'' package & & \\
$\PS$ & $\definedas$ & 
$\left\{\begin{array}{ll}
\P \st & \\
\quad |\P|=\algd, & \\
\quad \P \subseteq \I^*, & \\
\quad \P \preceq \I^* \backslash \P, & |\I^*| \geq \algd \\
\emptyset, & |\I^*| < \algd \\
\end{array}
\right.$ \\[2pt]
\end{tabular}

${}$

\noindent where $\algd \Z$ denotes integer multiples of $\algd$ and $\P \preceq
\I^* \backslash \P$ denotes that, for all $i \in \P$ and $j \in \I^*
\backslash \P$, $\mu_i \leq \mu_j$.  Then the following is a recursive
description of the algorithm:

${}$

\noindent \textbf{Recursive $\algd$-ary Package-Merge Procedure}

\textit{Basis.  $\rhot = 0$}:  $\os (\I,\rhot)=\emptyset$.

\textit{Case 1.  $\rho^* = \rhop$ and $\I \neq \emptyset$}:  $\os(\I,\rhot) =
\os (\I \backslash \{i^*\},\rhot-\rho^*) \cup \{i^*\}$.

\textit{Case 2a.  $\rho^* < \rhop$, $\I \neq \emptyset$, and $|\I^*| < \algd$}:
$\os(\I,\rhot) = \os(\I \backslash \I^*, \rhot)$.

\textit{Case 2b.  $\rho^* < \rhop$, $\I \neq \emptyset$, and $|\I^*| \geq \algd$}:
Create $i'$, a new item with weight $\mu_{i'} = \sum_{i\in \PS} \mu_i$
and width $\rho_{i'} = \algd \rho^*$.  This new item is thus a
combined item, or \defn{package}, formed by combining the $\algd$
least weighted items of width~$\rho^*$.  Let $\S = \os(\I \backslash \PS
\cup \{i'\},\rhot)$ (the optimization of the packaged version).  If $i' \in
\S$, then $\os(\I,\rhot) = \S \backslash \{i'\} \cup \PS$; otherwise,
$\os(\I,\rhot) = \S$.

\begin{figure*}[ht]
\centering
\psfrag{=0=0_3}{\large $\rhot=0=0_3$}
\psfrag{=3=10_3}{\large $\rhot=3=10_3$}
\psfrag{=5=12_3}{\large $\rhot=5=12_3$}
\psfrag{m=1}{\large $\mu=1$}
\psfrag{m=2}{\large $\mu=2$}
\psfrag{m=4}{\large $\mu=4$}
\psfrag{m=5}{\large $\mu=5$}
\psfrag{m=7}{\large $\mu=7$}
\psfrag{r=1}{\large $\rho=1$}
\psfrag{r=3}{\large $\rho=3$}
\psfrag{Sm=7}{\large $\sum \mu = 7$}
\resizebox{12cm}{!}{\includegraphics{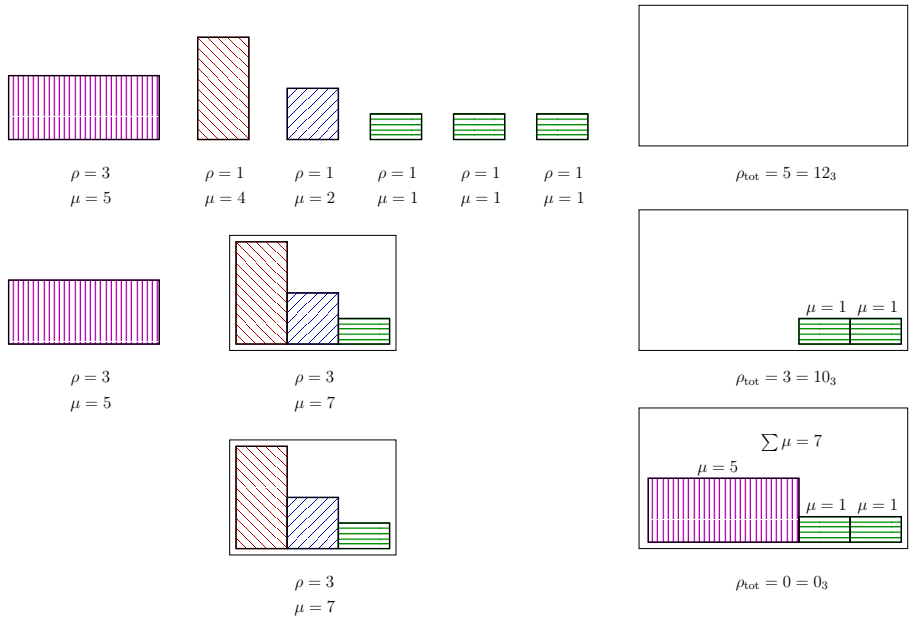}}
\caption{A simple example of the Package-Merge algorithm}
\label{pm}
\end{figure*}

\begin{theorem}
If an optimal solution to the Coin Collector's problem exists, the
above recursive (Package-Merge) algorithm will terminate with an
optimal solution. 
\end{theorem}

\begin{proof}
We show that the Package-Merge algorithm produces an optimal solution
via induction on the number of input items.  The basis is trivially correct,
and each inductive case reduces the number of items by at least one.
The inductive hypothesis on $\rhot \geq 0$ and $\I \neq \emptyset$ is
that the algorithm is correct for any problem instance with fewer
input items than instance $(\I,\rhot)$.

If $\rho^* > \rhop > 0$, or if $\I=\emptyset$ and $\rhot \neq 0$, then there
is no solution to the problem, contrary to our assumption.  Thus all
feasible cases are covered by those given in the procedure.  Case 1
indicates that the solution must contain at least one element (item or
package) of width~$\rho^*$.  These must include the minimum weight
item in $\I^*$, since otherwise we could substitute one of the items
with this ``first'' item and achieve improvement.  Case~2 indicates
that the solution must contain a number of elements of width $\rho^*$
that is a multiple of~$\algd$.  If this number is~$0$, none of the
items in $\PS$ are in the solution.  If it is not, then they all are.
Thus, if $\PS = \emptyset$, the number is~$0$, and we have Case 2a.
If not, we may ``package'' the items, considering the replaced package
as one item, as in Case 2b.  Thus the inductive hypothesis holds.
\end{proof}

Fig.~\ref{pm} presents a simple example of this algorithm at work
for $\algd=3$, finding minimum total weight items of total width
$\rhot=5$ (or, in ternary, $12_3$).  In the figure, item width
represents numeric width and item area represents numeric weight.
Initially, as shown in the top row, the minimum weight item has width
$\rho^* = \rho_{i^*} = \rhop = 1$.  This item is put into the solution
set, and the next step repeats the task on the items remaining outside
the solution set.  Then, the remaining minimum width items are
packaged into a merged item of width $3$ ($10_3$), as in the middle
row.  Finally, the minimum weight item/package with width $\rho^* =
\rho_{i^*} = \rhop = 3$ is added to complete the solution set, which
is now of weight~$7$.  The remaining packaged item is left out in this
case; when the algorithm is used for coding, several items are usually
left out of the optimal set.  Given input sorted first by width then
weight, the resulting algorithm is $\order(m)$ time and space.

\section{A General Algorithm}
\label{algorithm}

We now formalize the reduction from the coding problem to the Coin
Collector's problem.  This generalizes the similar reduction shown in
\cite{Baer06} for binary codes with only a limit on maximum length,
which is in turn a generalization of \cite{LaHi} for length-limited
binary codes with linear $\varphi$, the traditional penalty function.

We assert that any optimal solution $N$ of the Coin Collector's
problem with total width
$$\rhot=\frac{ n-\algd^{l_{\min}} } {\algd-1}\algd^{-l_{\min}}$$ on
coins $\I$ (identical to the set of all possible nodes $I$) is a
nodeset for an optimal solution of the coding problem.  This yields a
suitable method for solving the problem.

To show this reduction, we first define $\rho(N)$ in a natural manner
for any $N = \nodeset(\boldl)$:
\begin{eqnarray*}
\rho(N) &\definedas & \sum_{(i,l) \in N} \rho(i,l) \\
&=& \sum_{i=1}^n \sum_{l=l_{\min}+1}^{l_i} \algd^{-l} \\ 
&=& \sum_{i=1}^n \frac{\algd^{-l_{\min}}-\algd^{-l_i}}{\algd-1} \\ 
&=& \frac{n\algd^{-l_{\min}} - \letterk(\boldl)}{\algd-1}
\end{eqnarray*}
where $\letterk(\boldl)$ is the Kraft sum~(\ref{kraft}).  Given $n
\bmod{(\algd-1)} \equiv 1$, all optimal codes have the Kraft
inequality satisfied with equality; otherwise, the longest codeword
length could be shortened by one, strictly decreasing the penalty
without violating the inequality.  Thus the optimal solution has
$\letterk(\boldl)=1$ and
$$\rho(N)=\frac{n-\algd^{l_{\min}}}{\algd-1}\algd^{-l_{\min}}.$$

Also define:
\begin{eqnarray*}
\delta_\varphi(l) &\definedas& \varphi(l) - \varphi(l-1) \\
\mu(N) &\definedas & \sum_{(i,l) \in N} \mu(i,l).
\end{eqnarray*}
Note that
\begin{eqnarray*}
\mu(N) &=& \sum_{(i,l) \in N} \mu(i,l) \\
&=& \sum_{i=1}^n \sum_{l=l_{\min}+1}^{l_i} p_i \delta_\varphi(l-l_{\min}) \\
&=& \sum_{i=1}^n p_i \varphi(l_i-l_{\min}) - \sum_{i=1}^n p_i \varphi(0).
\end{eqnarray*}
Since the subtracted term is a constant, if the optimal nodeset
corresponds to a valid code, solving the Coin Collector's problem
solves this coding problem.  To prove the reduction, we need to prove
that the optimal nodeset indeed corresponds to a valid code.  We begin
with the following lemma:

\begin{lemma}
\label{lllemma}
Suppose that $N$ is a nodeset of width $x\algd^{-k}+r$ where $k$ and $x$ are
integers and $0<r<\algd^{-k}$.  Then $N$ has a subset $R$ with width~$r$.
\end{lemma}

\begin{proof}
  Let us use induction on the cardinality of the set.  The base case
  $|N|=1$ is trivial since then $x=0$.  Assume the lemma holds for all
  $|N| < n$, and suppose $|\tilde{N}|=n$.  Let $\rho^* = \min_{j \in
  \tilde{N}} \rho_j$ and $j^* = \argmin_{j \in \tilde{N}} \rho_j$.  We
  can view item $j^*$ of width $\rho^* \in \algd^\Z$ as the smallest
  contributor to the width of $\tilde{N}$ and $r$ as the portion of the
  $\algd$-ary expansion of the width of $\tilde{N}$ to the right 
  of~$\algd^{-k}$.  Then $r$ must be an integer multiple 
  of~$\rho^*$.  If $r=\rho^*$, $R =\{j^*\}$ is a solution.  Otherwise let
  $N'=\tilde{N} \backslash \{j^*\}$ (so $|N'|=n-1$) and let $R'$ be
  the subset obtained from solving the lemma for set $N'$ of width $r
  - \rho^*$.  Then $R=R'\cup \{j^*\}$.
\end{proof}

We now prove the reduction:

\begin{theorem}
\label{cceqll}
Any $N$ that is a solution of the Coin Collector's problem for $$\rhot
= \rho(N)=\frac{n-\algd^{l_{\min}}}{\algd-1}\algd^{-l_{\min}}$$ has a
corresponding length vector $\boldlupN$ such that $N =
\nodeset(\boldlupN)$ and $\mu(N) = \min_{\smalll} \sum_i p_i
\varphi(l_i-l_{\min}) - \varphi(0) \sum_i p_i$.
\end{theorem}

\begin{proof}
  Any optimal length vector nodeset has $\rho(\nodeset(\boldl))
  = \rhot$.  Suppose $N$ is a solution to the Coin Collector's problem but
  is not a valid nodeset of a length vector.  Then there exists
  an $(i,l)$ with $l \in [l_{\min}+2, l_{\max}]$ such that $(i,l) \in N$ and $(i,l-1)
  \in I \backslash N$.  Let $R' = N \cup \{(i,l-1)\} \backslash
  \{(i,l)\}$.  Then $\rho(R')=\rhot+(\algd-1)\algd^{-l}$ and, due to
  convexity, $\mu(R') \leq \mu(N)$.  Using $n
  \bmod{(\algd-1)} \equiv 1$, we know that $\rhot$ is an integer multiple
  of $\algd^{-l_{\min}}$.  Thus, using Lemma~\ref{lllemma} with
  $k=l_{\min}$, $x=\rhot\algd^{l_{\min}}$, and $r=(\algd-1)\algd^{-l}$,
  there exists an $R \subset R'$ such that $\rho(R)=r$.  Since
  $\mu(R)>0$, $\mu(R' \backslash R) < \mu(R') \leq
  \mu(N)$.  This is a contradiction to $N$ being an optimal
  solution to the Coin Collector's problem, 
  and thus any optimal solution of the Coin Collector's problem
  corresponds to an optimal length vector.
\end{proof}

Because the Coin Collector's problem is linear in time and space ---
same-width inputs are presorted by weight, numerical operations and
comparisons are constant time --- the overall algorithm finds an
optimal code in $\order(n(l_{\max}-l_{\min}))$ time and space.  Space
complexity, however, can be decreased.

\section{A Deterministic $\order(n)$-Space Algorithm}
\label{linear}

If $p_i = p_j$, we are guaranteed no particular inequality relation
between $l_i$ and $l_j$ since we did not specify a method for breaking
ties.  Thus the length vector returned by the algorithm need not
have the property that $l_i \leq l_j$ whenever $i < j$.  We would like
to have an algorithm that has such a monotonicity property.

\begin{definition} A \defn{monotonic} nodeset, $N$, is one with the following properties:
\begin{eqnarray}
&(i,l) \in N \Rightarrow (i+1,l) \in N& \for i<n \label{firstprop} \quad \\
&(i,l) \in N \Rightarrow (i,l-1) \in N& \for l>l_{\min}+1 . \quad \label{validlen} 
\end{eqnarray}
In other words, a nodeset is monotonic if and only if it corresponds
to a length vector $\boldl$ with lengths sorted in increasing
order; this definition is equivalent to that given in \cite{LaHi}.
\end{definition}

\begin{figure*}[ht]
\centering
\psfrag{A}{$\AlphaSet$}
\psfrag{B}{$\BetaSet$}
\psfrag{C}{$\GammaSet$}
\psfrag{D}{$\DeltaSet$}
\psfrag{N}{$N$}
\psfrag{3negmin}{$\algd^{-l_{\min}+1}$}
\psfrag{3negmid}{$\algd^{-l_{\mid}}$}
\psfrag{3negmax}{$\algd^{-l_{\max}}$}
\psfrag{lmin}{$l_{\min}+1$}
\psfrag{lmid}{$l_\mid$}
\psfrag{lmax}{$l_{\max}$}
\psfrag{n}{$n$}
\psfrag{n-m}{$n-n_\nu$}
\psfrag{1}{$1$}
\psfrag{l (level)}{$l$ (level)}
\psfrag{i (item)}{$i$ (input symbol)}
\psfrag{(width)}{$\rho$ (width)}
\resizebox{13cm}{!}{\includegraphics{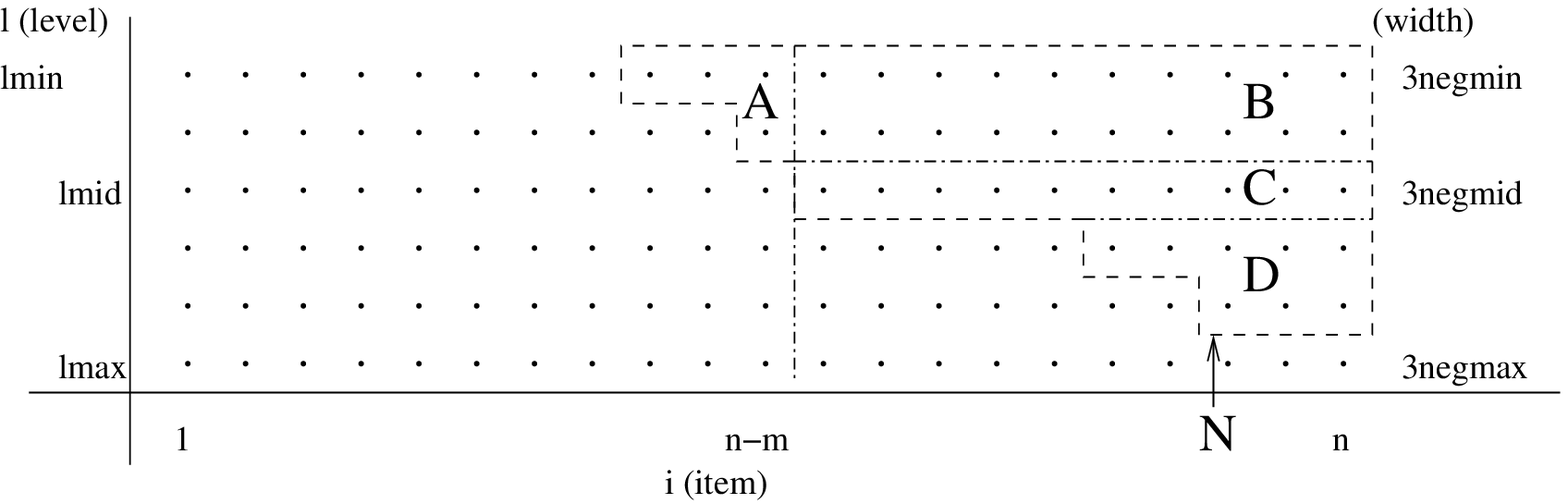}}
\caption{The set of nodes $I$, an optimal nodeset $N$, and disjoint subsets $\AlphaSet$, $\BetaSet$, $\GammaSet$, $\DeltaSet$}
\label{ABCD}
\end{figure*}

Examples of monotonic nodesets include the sets of nodes enclosed by
dashed lines in Fig.~\ref{nodesetnum} and Fig.~\ref{ABCD}.  In the
latter case, $n = 21$, $\algd = 3$, $l_{\min} = 2$, and $l_{\max} =
8$, so $\rhot = 2/3$.  As indicated, if $p_i = p_j$ for some $i$ and
$j$, then an optimal nodeset need not be monotonic.  However, if all
probabilities are distinct, the optimal nodeset is monotonic.

\begin{lemma}
\label{dmlemma}
If $\p$ has no repeated values, then any optimal solution
$N=\os(I,n-1)$ is monotonic.
\end{lemma}

\begin{proof}
The second monotonic property (\ref{validlen}) was proved for optimal
nodesets in Theorem~\ref{cceqll}.  The first property
(\ref{firstprop}) can be shown via a simple exchange argument.
Consider optimal $\boldl$ with $i>j$ so that $p_i < p_j$, and also
consider $\boldl'$ with lengths for inputs $i$ and $j$ interchanged,
as in \cite[pp.~97--98]{CoTh}.  Then
$$
\begin{array}{l}
\sum_k p_k \varphi(l'_k-l_{\min}) - \sum_k p_k \varphi(l_k-l_{\min}) \\
\quad = (p_j - p_i)\left[\varphi(l_i-l_{\min}) - 
 \varphi(l_j-l_{\min})\right] \\
\quad \leq 0 
\end{array}
$$ where the inequality is to due to the optimality of~$\boldl$.
Since $p_j-p_i > 0$ and $\varphi$ is monotonically increasing, $l_i
\geq l_j$ for all $i>j$ and an optimal nodeset without repeated $\p$
must be monotonic.
\end{proof}

Taking advantage of monotonicity in a Package-Merge coding
implementation to trade off a constant factor of time for drastically
reduced space complexity is done in \cite{Larm} for length-limited
binary codes.  We extend this to the length-bounded problem, first for
$\p$ without repeated values, then for arbitrary~$\p$.

Note that the total width of items that are each less than or equal to
width $\rho$ is less than~$2n\rho$.  Thus, when we are processing
items and packages of width $\rho$, fewer than $2n$ packages are
kept in memory.  The key idea in reducing space complexity is to keep
only four attributes of each package in memory instead of the full
contents.  In this manner, we use $\order(n)$ space while retaining enough
information to reconstruct the optimal nodeset in algorithmic
postprocessing.

Define $$l_{\mid} \definedas \left\lfloor \frac{1}{2}
(l_{\max}+l_{\min}+1) \right\rfloor .$$ For each package $S$, we
retain only the following attributes:
\begin{enumerate}
\item $\mu(S) \das \sum_{(i,l) \in S} \mu(i,l)$
\item $\rho(S) \das \sum_{(i,l) \in S} \rho(i,l)$
\item $\nu(S) \das |S \cap I_{\mid}|$
\item $\psi(S) \das \sum_{(i,l) \in S \cap I_{\hi}} \rho(i,l)$
\end{enumerate}
where $I_{\hi} \definedas \{ (i,l)~|~l>l_{\mid} \}$ and $I_{\mid}
\das \{ (i,l)~|~l=l_{\mid} \}$.  We also define $I_\lo \das
\{(i,l)~|~l<l_{\mid} \}$.

With only these parameters, the ``first run'' of the algorithm takes
$\order(n)$ space.  The output of this run is the package attributes
of the optimal nodeset $N$.  Thus, at the end of this first run, we
know the value for $n_\nu \das \nu(N)$, and we can consider $N$ as the
disjoint union of four sets, shown in Fig.~\ref{ABCD}:
\begin{enumerate}
\item $\AlphaSet$ = nodes in $N \cap I_\lo$ with indices in $[1,n-n_\nu]$,
\item $\BetaSet$ = nodes in $N \cap I_\lo$ with indices in $[n-n_\nu+1,n]$,
\item $\GammaSet$ = nodes in $N \cap I_{\mid}$,
\item $\DeltaSet$ = nodes in $N \cap I_\hi$.
\end{enumerate}
Due to the monotonicity of $N$, it is clear that $\BetaSet =
[n-n_\nu+1,n] \times [l_{\min}+1, l_\mid-1]$ and $\GammaSet =
[n-n_\nu+1, n] \times \{l_\mid\}$.  Note then that $\rho(\BetaSet) =
(n_\nu) (\algd^{-l_{\min}}-\algd^{1-l_{\mid}})/(\algd-1)$ and
$\rho(\GammaSet) = n_\nu \algd^{-l_\mid}$.  Thus we need merely to
recompute which nodes are in $\AlphaSet$ and in~$\DeltaSet$.

Because $\DeltaSet$ is a subset of $I_{\hi}$, $\rho(\DeltaSet) =
\psi(N)$ and $\rho(\AlphaSet) = \rho(N) - \rho(\BetaSet) -
\rho(\GammaSet) - \rho(\DeltaSet)$.  Given their respective widths,
$\AlphaSet$ is a minimal weight subset of $[1,n-n_\nu] \times
[l_{\min}+1,l_{\mid}-1]$ and $\DeltaSet$ is a minimal weight subset of
$[n-n_\nu+1, n] \times [l_{\mid}+1,l_{\max}]$.  These are
monotonic if the overall nodeset is monotonic.  The nodes at each
level of $\AlphaSet$ and $\DeltaSet$ can thus be found by recursive
calls to the algorithm.  This approach uses only $\order(n)$ space
while preserving time complexity; one run of an algorithm on
$n(l_{\max}-l_{\min})$ nodes is replaced with a series of runs, first
one on $n(l_{\max}-l_{\min})$ nodes, then two on an average of at most
$n(l_{\max}-l_{\min})/4$ nodes each, then four on an average of at
most $n(l_{\max}-l_{\min})/16$, and so forth.  An optimization of
the same complexity is made in \cite{LaHi}, where it is proven that
this yields $\order(n(l_{\max}-l_{\min}))$ time complexity with a
linear space requirement.  Given the hard bounds for $l_{\max}$ and
$l_{\min}$, this is always $\order(n^2/\algd)$.

The assumption of distinct $p_i\s$ puts an undesirable restriction on
our input that we now relax.  In doing so, we make the algorithm
deterministic, resolving ties that make certain minimization steps of
the algorithm implementation dependent.  This results in what in some sense is
the ``best'' optimal code if multiple monotonic optimal codes exist.

Recall that $\p$ is a nonincreasing vector.  Thus items of a given
width are sorted for use in the Package-Merge algorithm; this order is
used to break ties.  For example, if we look at the problem in
Fig.~\ref{nodesetnum} --- $\varphi(\delta) = \delta^2$, $n=7$,
$\algd=3$, $l_{\min} = 1$, $l_{\max}=4$ --- with probability vector
$\p = (0.4, 0.3, 0.14, 0.06, 0.06, 0.02, 0.02)$, then nodes
$(7,4)$, $(6,4)$, and $(5,4)$ are the first to be grouped, the tie
between $(5,4)$ and $(4,4)$ broken by order.  Thus, at any step, all
identical-width items in one package have adjacent indices.  Recall
that packages of items will be either in the final nodeset or absent
from it as a whole.  This scheme then prevents any of the
nonmonotonicity that identical $p_i\s$ might bring about.

In order to assure that the algorithm is fully deterministic, the
manner in which packages and single items are merged must also be
taken into account.  We choose to combine nonmerged items before merged
items in the case of ties, in a similar manner to the two-queue
bottom-merge method of Huffman coding\cite{Schw,Leeu}.  Thus, in our
example, there is a point at which the node $(2,2)$ is chosen (to be
merged with $(3,2)$ and $(4,2)$) while the identical-weight package of
items $(5,3)$, $(6,3)$, and $(7,3)$ is not.  This leads to the optimal
length vector $\boldl = (1, 2, 2, 2, 2, 2, 2)$, rather than $\boldl =
(1, 1, 2, 2, 3, 3, 3)$ or $\boldl = (1, 1, 2, 3, 2, 3, 3)$, which are
also optimal.  The corresponding nodeset is enclosed within the dashed
line in Fig.~\ref{nodesetnum}, and the resulting monotonic code tree
is the code tree shown in Fig.~\ref{codetree}.

This approach also enables us to set $\epsilon$, the value for dummy
variables, equal to $0$ without violating monotonicity.  As in
bottom-merge Huffman coding, the code with the minimum reverse
lexicographical order among optimal codes (and thus the one with
minimum height) is the one produced; reverse lexicographical order
is the lexicographical order of lengths after their being sorted
largest to smallest.  An identical result can be obtained by using
the position of the ``largest'' node in a package (in terms of 
position number $nl+i$) in order to choose those with lower values, as in
\cite{LaPr2}.  However, our approach, which can be shown to be
equivalent via simple induction, eliminates the need for keeping track
of the maximum value of $nl+i$ for each package.

\section{Further Refinements}
\label{refine}

There are changes we can make to the algorithm that, for certain
inputs, result in even better performance.  For example, if $l_{\max}
\approx \log_\algd n$, then, rather than minimizing the weight of
nodes of a certain total width, it is easier to maximize weight over a
complementary total width and find the complementary set of nodes.
Similarly, if most input symbols have one of a handful of probability
values, one can consider this and simplify calculations.  These and
other similar optimizations have been done in the past for the special
case $\varphi(\delta)=\delta$, $l_{\min}=0$,
$\algd=2$\cite{KMT,LiMo,MTK,TuMo,TuMo2}, though we do not address or
extend such improvements here.

So far we have assumed that $l_{\max}$ is the best upper bound on
codeword length we could obtain.  However, there are many cases in
which we can narrow the range of codeword lengths, thus making the
algorithm faster.  For example, since, as stated previously, we can
assume without loss of generality that $l_{\max} \leq \lceil
(n-1)/(\algd-1) \rceil$, we can eliminate the bottom row of nodes
from consideration in Fig.~\ref{nodesetnum}.

Consider also when $l_{\min}=0$.  An upper bound on $\{l_i\}$ can
be derived from a theorem and a definition due to Larmore:

\begin{definition} 
Consider penalty functions $\varphi$ and~$\chi$.  We say that $\chi$
is \defn{flatter} than $\varphi$ if, for positive integers $l' > l$,
$(\chi(l)-\chi(l-1))(\varphi(l')-\varphi(l'-1)) \leq
(\varphi(l)-\varphi(l-1))(\chi(l')-\chi(l'-1))$.
\cite{Larm}.
\end{definition} 

A consequence of the Convex Hull Theorem of \cite{Larm} is that, given
$\chi$ flatter than $\varphi$, for any $\p$, there exist
$\varphi$-optimal $\boldlvarphi$ and $\chi$-optimal
$\boldlchi$ such that $\boldlvarphi$ is greater
than $\boldlchi$ in terms of reverse lexicographical order.
This explains why the word ``flatter'' is used.

Penalties flatter than the linear penalty --- i.e., convex~$\varphi$
--- can therefore yield a useful upper bound, reducing complexity.
Thus, if $l_{\min}=0$, we can use the results of a pre-algorithmic
Huffman coding of the input symbols to find an upper bound on codeword
length in linear time, one that might be better than~$l_{\max}$.
Alternatively, we can use the least probable input to find a looser
upper bound, as in \cite{CaDe2}.

${}$

When $l_{\min}>1$, one can still use a modified pre-algorithmic
Huffman coding to find an upper bound as long as
$\varphi(\delta)=\delta$.  This is done via a modification of the
Huffman algorithm allowing an arbitrary minimum $l_{\min}$ and a
trivial maximum (e.g., $l_{\max} = n$ or $\lceil (n-1)/(\algd-1)
\rceil$):

\begin{center}
{\bf Procedure for length-lower-bounded (``truncated Huffman'') coding}
\end{center}

\begin{enumerate}
\item Add $(\algd-n) \bmod{(\algd-1)}$ dummy items of probability~$0$.
\item Combine the items with the $\algd$ smallest probabilities
  $p_{i_1}, p_{i_2}, \ldots, p_{i_\algd}$ into one item with the
  combined probability $\tilde{p}_i = \sum_{i=1}^\algd p_{i_1}$.  This item
  has codeword $\tilde{c}_i$, to be determined later, while these
  $\algd$ smallest items are assigned concatenations of this
  yet-to-be-determined codeword and every possible output symbol, that
  is, $c_{i_1} = \tilde{c}_i0, c_{i_2} = \tilde{c}_i1, \ldots, c_{i_D}
  = \tilde{c}_i (\algd-1)$.  Since these have been assigned in terms
  of $\tilde{c}_i$, replace the smallest $\algd$ items with
  $\tilde{p}_i$ in $\p$ to form $\tilde{\p}$.
\item Repeat previous step, now with the remaining $n-\algd+1$
  codewords and corresponding probabilities, until only
  $\algd^{l_{\min}}$ items are left.  
\item Assign all possible $l_{\min}$ long codewords to these items,
thus defining the overall code based on the fixed-length code assigned
to these combined items.
\end{enumerate}

This procedure is Huffman coding truncated midway through coding,
the resulting trees serving as subtrees of nodes of identical depth.
Excluding the last step, the algorithm is identical to that 
shown in \cite{Tome} to result in an optimal Huffman forest.  The
optimality of the algorithm for length-lower-bounded coding is an
immediate consequence of the optimality of the forest, as both have
the same constraints and the same value to minimize.  As with the
usual Huffman algorithm, this can be made linear time given sorted
inputs\cite{Leeu} and can be made to find a code with the minimum
reverse lexicographical order among optimal codes via the bottom-merge
variant.

Clearly, this algorithm finds the optimal code for the length-bounded
problem if the resulting code has no codeword longer than $l_{\max}$,
whether this be because $l_{\max}$ is trivial or because of other
specifications of the problem.  If this truncated Huffman algorithm
fails, then we know that $l_n = l_{\max}$, that is, we cannot have
that $l_n < l_{\max}$ for the length-bounded code.  This is an
intuitive result, but one worth stating and proving, as it is
used in the next section:

\begin{lemma}
\label{hclemma}
If a (truncated) Huffman code ($\varphi(\delta) = \delta$) for
$l_{\min}$ has a codeword longer than some $l_{\ub}$, then there exists an
optimal length-bounded code for bound $[l_{\min}, l_{\ub}]$ with
codewords of length $l_{\ub}$.
\end{lemma}

\begin{proof}
It suffices to show that, if an optimal code for the bound $[l_{\min},
l_{\max}]$ has a codeword with length $l_{\max}$, then an optimal code
for the bound $[l_{\min}, l_{\max}-1]$ has a codeword with length
$l_{\max}-1$, since this can be applied inductively from $l_{\max} =
l_n$ (assuming $l_n$ is the length of the longest codeword of the
truncated Huffman code) to $l_{\ub}$, obtaining the desired result.
The optimal nodeset $N$ for the bound $[l_{\min}, l_{\max}]$ has width
$\algd^{-l_{\min}}(n-\algd^{l_{\min}})/(\algd-1)$.  Therefore, in the
course of the Package-Merge algorithm, we at one point have
$(n-\algd^{l_{\min}})/(\algd-1)$ packages of width $\algd^{-l_{\min}}$
which will eventually comprise optimal nodeset $N$, these packages
having weight no larger than the remaining packages of the same width.

Consider the nodeset $N'$ formed by making each $(i,l)$ in $N$ into
$(i,l-1)$.  This nodeset is the solution to the Package-Merge
algorithm for the total width
$\algd^{-{l_{\min}}+1}(n-\algd^{l_{\min}})/(\algd-1)$ with bounds
$l_{\min}-1$ and $l_{\max}-1$.  Let $i(l)$ denote the number of nodes
on level $l$.  Then $i(l_{\min}) \geq n - \algd^{l_{\min}}$ since at
most $\algd^{l_{\min}}$ nodes can have length $l_{\min}$.  The subset of
$N'$ not of depth $l_{\min}-1$ is thus an optimal solution for bounds
$l_{\min}$ and $l_{\max}-1$ with total width 
$$\algd^{-l_{\min}}\left(\frac{\algd(n-\algd^{l_{\min}})}{\algd-1}-\algd
i(l_{\min})\right)$$ that is, at one point in the algorithm this
solution corresponds to the $\algd
(n-\algd^{l_{\min}})/(\algd-1)-i(l_{\min})$ least weighted packages of
width $\algd^{-l_{\min}}$.  Due to the bounds on $i(l_{\min})$, this
number of packages is less than the number of packages of the same
width in the optimal nodeset for bounds $l_{\min}$ and $l_{\max}-1$
(with total width $\algd^{l_{\min}}(n-\algd^{l_{\min}})/(\algd-1)$).
Thus an optimal nodeset to the shortened problem can contain the
(shifted-by-one) original nodeset and must have its maximum length
achieved for all input symbols for which the original nodeset achieves
maximum length.
\end{proof}

Thus we can find whether $l_n=l_{\max}$ by merely doing
pre-algorithmic bottom-merge Huffman coding (which, when $l_n \neq
l_{\max}$, results in reduced computation).  This is useful in
finding a faster algorithm for large $l_{\max}-l_{\min}$ and
linear~$\varphi$.

\section{A Faster Algorithm for the Linear Penalty}
\label{linpen}

A somewhat different reduction, one analogous to the reduction of
\cite{LaPr1}, is applicable if $\varphi(\delta) = \delta$.  This more
specific algorithm has similar space complexity and strictly better
time complexity unless $l_{\max}-l_{\min} = \order(\log n)$.  However, we
only sketch this approach here roughly compared to our previous
explanation of the simpler, more general approach.

Consider again the code tree representation, that using a $\algd$-ary
tree to represent the code.  A codeword is represented by successive
splits from the root to a leaf --- one split for each output symbol
--- so that the length of a codeword is represented by the length of
the path to its corresponding leaf.  A vertex that is not a leaf is
called an \defn{internal vertex}; each internal vertex of the tree in
Fig.~\ref{codetree} is shown as a black circle.  We continue to use
dummy variables to ensure that $n \bmod{(\algd-1)} \equiv 1$, and thus
an optimal tree has $\letterk(\boldl)=1$; equivalently, all internal
vertices have $\algd$ children.  We also continue to assume without
loss of generality that the output tree is monotonic.  An optimal tree
given the constraints of the problem will have no internal vertices at
level $l_{\max}$, $(n-\algd^{l_{\min}})/(\algd-1)$ internal vertices
in the $l_{\max}-l_{\min}$ previous levels, and
$(\algd^{l_{\min}}-1)/(\algd-1)$ internal vertices --- with no leaves
--- in the levels above this, if any.  The solution to a linear
length-bounded problem can be expressed by the number of internal
vertices in the unknown levels, that is, by
\begin{equation}
\begin{array}{rcl}
\alpha_i &\definedas& \mbox{number of internal vertices} \\
&& \mbox{in levels } [l_{\max}-i,l_{\max}]
\end{array}
\label{alpha}
\end{equation}
so that we know that $$\alpha_0 = 0 \quad \mbox{and} \quad
\alpha_{l_{\max}-l_{\min}} = \frac{n-\algd^{l_{\min}}}{\algd-1}.$$

\begin{figure}[ht]
\centering
\psfrag{p27}{\tiny $p_2+p_3+p_4+p_5+p_6+p_7$}
\psfrag{p37}{\tiny $p_3+p_4+p_5+p_6+p_7$}
\psfrag{p57}{\tiny $p_5+p_6+p_7$}
\includegraphics{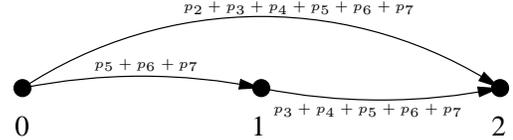}
\caption{The directed acyclic graph for coding $n=7$, $\algd=3$,
$l_{\min} = 1$, $l_{\max}=l_n=4$ ($\varphi(\delta) = \delta$)}
\label{DAG}
\end{figure}

If the truncated Huffman coding algorithm (as in the previous section)
fails to find a code with all $l_i \leq l_{\max}$, then we are assured
that there exists an $l_i = l_{\max}$, so that $\alpha_i$ can be
assumed to be a sequence of strictly increasing integers.  A strictly
increasing sequence can be represented by a path on a different type
of graph, a directed acyclic graph with vertices numbered $0$ to
$(n-\algd^{l_{\min}})/(\algd-1)$, e.g., the graph of vertices in
Fig.~\ref{DAG}.  The $i$th edge of the path begins at $\alpha_{i-1}$
and ends at $\alpha_i$, and each $\alpha_i$ represents the number of
internal vertices at and below the corresponding level of the tree
according to (\ref{alpha}).  Fig.~\ref{codetree} shows a code tree
with corresponding $\alpha_i\s$ as a count of internal vertices.  The
path length is identical to the height of the corresponding tree, and
the path weight is
$$\sum_{i=1}^{l_{\max}-l_{\min}} w(\alpha_{i-1}, \alpha_i)$$ for edge
weight function $w$, to be determined.  Larmore and Przytycka used
such a representation for binary codes\cite{LaPr1}; here we use the
generalized representation for $\algd$-ary codes.

In order to make this representation correspond to the above problem,
we need a way of making weighted path length correspond to coding
penalty and a way of assuring a one-to-one correspondence between
valid paths and valid monotonic code trees.  First let us define the
cumulative probabilities
$$s_i \definedas \sum_{k=n-i+1}^n p_k$$ so that there are $n+1$
possible values for $s_i$, each of which can be accessed in constant
time after $\order(n)$-time preprocessing.  We then use these values
to weigh paths such that
$$
w(\alpha', \alpha'') \definedas \left\{\begin{array}{ll}
s_{(\algd \alpha''^+ - \alpha'^+)} ,&\algd \alpha''^+ - \alpha'^+ \leq n\\
\posinf,&\algd \alpha''^+ - \alpha'^+ > n\\
\end{array}
\right. 
$$ where we recall that $x^+$ denotes $\max(x,0)$ and $\posinf$ is
necessary for cases in which the numbers of internal vertices are
incompatible; this rules out paths not corresponding to valid trees.
Thus path length and penalty are equal, that is,
$$\sum_{i=1}^{l_{\max}-l_{\min}} w(\alpha_{i-1}, \alpha_i) =
\sum_{j=1}^n p_j (l_j - l_{\min}).$$

This graph weighting has the \defn{concave Monge property} or \defn{quadrangle
inequality}, 
$$
\begin{array}{l}
w(\alpha',\alpha'') + w(\alpha'+1,\alpha''+1) \\
\quad \leq w(\alpha',\alpha''+1)+w(\alpha'+1,\alpha'')
\end{array}
$$
for all $0<\alpha'+1<\alpha''\leq(n-\algd^{l_{\min}})/(\algd-1)$,
since this inequality reduces to the already-assumed $p_{n-\algd
\alpha'' + \alpha' + 1 - \algd} \geq p_{n-\algd \alpha'' + \alpha' +
2}$ (where $p_i \das 0$ for $i>n$).  Fig.~\ref{DAG} shows such a
graph.  A single-edge path corresponds to $\boldl = (1, 2, 2, 2, 2, 2,
2)$ while the two-edge path corresponds to $\boldl = (1, 1, 2, 2, 3,
3, 3)$.  In practice, only the latter would be under consideration
using the algorithm in question, since the pre-algorithmic Huffman
coding assured that $l_n = l_{\max} = 3$. 

Thus, if $$k \definedas l_{\max}-l_{\min}$$ and $$n' \definedas 1 +
\frac{n-\algd^{l_{\min}}}{\algd-1}$$ we wish to find the minimum
$k$-link path from $0$ to $(n-\algd^{l_{\min}})/(\algd-1)$ on this
weighted graph of $n'$ vertices.  Given the concave Monge property, an
$n'2^{\order(\sqrt{\log k \log \log n'})}$-time $\order(n')$-space
algorithm for solving this problem is presented in \cite{Schi}.  Thus
the problem in question can be solved in $n2^{\order(\sqrt{\log
(l_{\max}-l_{\min}) \log \log n})}/\algd$ time and $\order(n/\algd)$
space --- $\order(n)$ space if one counts the pre-algorithmic Huffman
coding and/or necessary reconstruction of the Huffman code or codeword
lengths --- an improvement on the Package-Merge-based approach except
for $k=\order(\log n)$.

\section{Extensions}
\label{conclusion}

One might wonder whether the time complexity of the aforementioned
algorithms is the minimum achievable.  
Special cases (e.g., $l_{\max} \approx \log_\algd n$ for
$\varphi(\delta)=\delta$, $l_{\min}=0$, and $\algd=2$) can be
addressed using modifications of the Package-Merge
approach\cite{KMT,LiMo,MTK,TuMo,TuMo2}.  Also, $\p$ often implies
ranges of values, obtainable without coding, for $l_1$ and~$l_n$.
This enables one to use values of $l_{\min}$ and $l_{\max}$ that
result in a significant improvement, as in \cite{Baer06} for $l_{\min}
= 0$.

An important problem that can be solved with the techniques in this paper
is that of finding an optimal code given an upper bound on fringe, the
difference between minimum and maximum codeword length.  One might,
for example, wish to find a fringe-limited prefix code in order to
have a near-optimal code that can be simply implemented, as in Section
VIII of~\cite{KEGS}.  Such a problem is mentioned in
\cite[p.~121]{Abr01}, where it is suggested that if there are $b-1$ codes
better than the best code having fringe at most $d$, one can find this
$b$-best code with the $\order(b n^3)$-time algorithm in
\cite[pp.~890--891]{AnHa}, thus solving the fringe-limited problem.
However, this presumes we know an upper bound for $b$ before running
this algorithm.  More importantly, if a probability vector is far from
uniform, $b$ can be very large, since the number of viable code trees
is $\Theta(1.794\ldots^n)$\cite{KMN, FlPr}.  Thus this is a poor
approach in general.  

Instead, we can use the aforementioned algorithms for finding the
optimal length-bounded code with codeword lengths restricted to
$[l'-d,l']$ for each $l' \in \{\lceil \log_\algd n \rceil, \lceil
\log_\algd n \rceil + 1, \ldots, \lfloor \log_\algd n \rfloor + d\}$,
keeping the best of these codes; this covers all feasible cases of
fringe upper bounded by~$d$.  (Here we again assume, without loss of
generality, that $n \bmod{(\algd-1)} \equiv 1$.)  The overall
procedure thus has time complexity $\order(nd^2)$ for the general
convex quasiarithmetic case and $nd2^{\order(\sqrt{\log d \log \log
n})}/\algd$ when applying the algorithm of Section~\ref{linpen} to the
most common penalty of expected length; the latter approach is of
lower complexity unless $d=\order(\log n)$.  Both algorithms operate
with only $\order(n)$ space complexity.

\section*{Acknowledgments}

The author wishes to thank Zhen Zhang for first bringing a related
problem to his attention and David Morgenthaler for constructive
discussions on this topic.

\bibliographystyle{IEEEtranS}

\begin{thebibliography}{10}
\providecommand{\url}[1]{#1}
\csname url@rmstyle\endcsname
\providecommand{\newblock}{\relax}
\providecommand{\bibinfo}[2]{#2}
\providecommand\BIBentrySTDinterwordspacing{\spaceskip=0pt\relax}
\providecommand\BIBentryALTinterwordstretchfactor{4}
\providecommand\BIBentryALTinterwordspacing{\spaceskip=\fontdimen2\font plus
\BIBentryALTinterwordstretchfactor\fontdimen3\font minus
  \fontdimen4\font\relax}
\providecommand\BIBforeignlanguage[2]{{%
\expandafter\ifx\csname l@#1\endcsname\relax
\typeout{** WARNING: IEEEtran.bst: No hyphenation pattern has been}%
\typeout{** loaded for the language `#1'. Using the pattern for}%
\typeout{** the default language instead.}%
\else
\language=\csname l@#1\endcsname
\fi
#2}}

\bibitem{Abr01}
J.~Abrahams, ``Code and parse trees for lossless source encoding,''
  \emph{Communications in Information and Systems}, vol.~1, no.~2, pp.
  113--146, Apr. 2001.

\bibitem{AnHa}
S.~Anily and R.~Hassin, ``Ranking the best binary trees,'' \emph{SIAM J.
  Comput.}, vol.~18, no.~5, pp. 882--892, Oct. 1989.

\bibitem{Baer06}
M.~B. Baer, ``Source coding for quasiarithmetic penalties,'' \emph{IEEE Trans.
  Inf. Theory}, vol. IT-52, no.~10, pp. 4380--4393, Oct. 2006.

\bibitem{CaDe2}
R.~M. Capocelli and A.~{De Santis}, ``A note on {$D$-ary Huffman} codes,''
  \emph{IEEE Trans. Inf. Theory}, vol. IT-37, no.~1, pp. 174--179, Jan. 1991.

\bibitem{Cass}
S.~Cass, ``Holiday gifts,'' \emph{IEEE Spectrum}, vol.~42, no.~11, pp. 59--68,
  Nov. 1994, available from \url{http://www.spectrum.ieee.org/nov05/2133/3}.

\bibitem{ChGo}
S.-L. Chan and M.~J. Golin, ``A dynamic programming algorithm for constructing
  optimal ``1''-ended binary prefix-free codes,'' \emph{IEEE Trans. Inf.
  Theory}, vol. IT-46, no.~4, pp. 1637--1644, July 2000.

\bibitem{ChTh}
C.~Chang and J.~Thomas, ``{Huffman} algebras for independent random
  variables,'' \emph{Disc. Event Dynamic Syst.}, vol.~4, no.~1, pp. 23--40,
  Feb. 1994.

\bibitem{CoTh}
T.~M. Cover and J.~A. Thomas, \emph{Elements of Information Theory},
  1st~ed.\hskip 1em plus 0.5em minus 0.4em\relax New York, NY:
  Wiley-Interscience, 1991.

\bibitem{DePe}
A.~{De Santis} and G.~Persiano, ``An optimal algorithm for the construction of
  optimal prefix codes with given fringe,'' in \emph{Proc., IEEE Data
  Compression Conf.}, Apr. 8--11, 1991, pp. 297--306.

\bibitem{Dic}
C.~Dickens, \emph{A Christmas Carol}.\hskip 1em plus 0.5em minus 0.4em\relax
  London, UK: Chapman and Hall, 1843, available from
  \url{http://www.gutenberg.org/etext/46}.

\bibitem{FlPr}
P.~Flajolet and H.~Prodinger, ``Level number sequences for trees,'' \emph{Disc.
  Math.}, vol.~65, no.~2, pp. 149--156, June 1987.

\bibitem{Gare}
M.~R. Garey, ``Optimal binary search trees with restricted maximal depth,''
  \emph{SIAM J. Comput.}, vol.~3, no.~2, pp. 101--110, June 1974.

\bibitem{GaWa}
A.~M. Garsia and M.~L. Wachs, ``A new algorithm for minimum cost binary
  trees,'' \emph{SIAM J. Comput.}, vol.~6, no.~4, pp. 622--642, Dec. 1977.

\bibitem{GoRo}
M.~J. Golin and G.~Rote, ``A dynamic programming algorithm for constructing
  optimal prefix-free codes for unequal letter costs,'' \emph{IEEE Trans. Inf.
  Theory}, vol. IT-44, no.~5, pp. 1770--1781, Sept. 1998.

\bibitem{GoWo}
L.~Gotlieb and D.~Wood, ``The construction of optimal multiway search trees and
  the monotonicity principle,'' \emph{Intern. J. Computer Maths, Section A},
  vol.~9, no.~1, pp. 17--24, 1981.

\bibitem{HKT}
T.~C. Hu, D.~J. Kleitman, and J.~K. Tamaki, ``Binary trees optimum under
  various criteria,'' \emph{SIAM J. Appl. Math.}, vol.~37, no.~2, pp. 246--256,
  Apr. 1979.

\bibitem{HLM}
T.~C. Hu, L.~L. Larmore, and J.~D. Morgenthaler, ``Optimal integer alphabetic
  trees in linear time,'' in \emph{Proc. 13th Annual European Symposium on
  Algorithms}.\hskip 1em plus 0.5em minus 0.4em\relax Springer-Verlag, Oct.
  2005, pp. 226--237.

\bibitem{HuMo}
T.~C. Hu and J.~D. Morgenthaler, ``Optimum alphabetic binary trees,'' in
  \emph{Combinatorics and Computer Science}, ser. Lecture Notes in Computer
  Science, vol. 1120.\hskip 1em plus 0.5em minus 0.4em\relax Springer-Verlag,
  Aug. 1996, pp. 234--243.

\bibitem{HuTa}
T.~C. Hu and K.~C. Tan, ``Path length of binary search trees,'' \emph{SIAM J.
  Appl. Math.}, vol.~22, no.~2, pp. 225--234, Mar. 1972.

\bibitem{HuTu}
T.~C. Hu and A.~C. Tucker, ``Optimal computer search trees and variable-length
  alphabetic codes,'' \emph{SIAM J. Appl. Math.}, vol.~21, no.~4, pp. 514--532,
  Dec. 1971.

\bibitem{Huff}
D.~A. Huffman, ``A method for the construction of minimum-redundancy codes,''
  \emph{Proc. IRE}, vol.~40, no.~9, pp. 1098--1101, Sept. 1952.

\bibitem{Humb2}
P.~A. Humblet, ``Generalization of {Huffman} coding to minimize the probability
  of buffer overflow,'' \emph{IEEE Trans. Inf. Theory}, vol. IT-27, no.~2, pp.
  230--232, Mar. 1981.

\bibitem{Itai}
A.~Itai, ``Optimal alphabetic trees,'' \emph{SIAM J. Comput.}, vol.~5, no.~1,
  pp. 9--18, Mar. 1976.

\bibitem{KMT}
J.~Katajainen, A.~Moffat, and A.~Turpin, ``A fast and space-economical
  algorithm for length-limited coding,'' in \emph{Proc., Int. Symp. on
  Algorithms and Computation}, Dec. 1995, p. 1221.

\bibitem{KEGS}
M.~Khosravifard, M.~Esmaeili, T.~A. Gulliver, and H.~Saidi, ``The minimum
  average code for finite memoryless monotone sources,'' in \emph{Proc., IEEE
  Information Theory Workshop}, Oct. 2002, pp. 135--138.

\bibitem{Knu71}
D.~E. Knuth, ``Optimum binary search trees,'' \emph{Acta Informatica}, vol.~1,
  pp. 14--25, 1971.

\bibitem{Knu31}
------, \emph{The Art of Computer Programming, Vol. 3: Sorting and Searching},
  1st~ed.\hskip 1em plus 0.5em minus 0.4em\relax Reading, MA: Addison-Wesley,
  1973.

\bibitem{KMN}
J.~Komlos, W.~Moser, and T.~Nemetz, ``On the asymptotic number of prefix
  codes,'' \emph{Mitteilungen aus dem Mathematischen Seminar Giessen, {\em Heft
  165, Coxeter Festschrift, Teil III}}, pp. 35--48, 1984.

\bibitem{Larm}
L.~L. Larmore, ``Minimum delay codes,'' \emph{SIAM J. Comput.}, vol.~18, no.~1,
  pp. 82--94, Feb. 1989.

\bibitem{LaHi}
L.~L. Larmore and D.~S. Hirschberg, ``A fast algorithm for optimal
  length-limited {Huffman} codes,'' \emph{J. ACM}, vol.~37, no.~2, pp.
  464--473, Apr. 1990.

\bibitem{LaPr1}
L.~L. Larmore and T.~M. Przytycka, ``Parallel construction of trees with
  optimal weighted path length,'' in \emph{Proc. 3nd Annual Symposium on
  Parallel Algorithms and Architectures}, 1991, pp. 71--80.

\bibitem{LaPr2}
------, ``A fast algorithm for optimal height-limited alphabetic
  binary-trees,'' \emph{SIAM J. Comput.}, vol.~23, no.~6, pp. 1283--1312, Dec.
  1994.

\bibitem{LiMo}
M.~Liddell and A.~Moffat, ``Incremental calculation of optimal
  length-restricted codes,'' in \emph{Proc., IEEE Data Compression Conf.}, Apr.
  2--4, 2002, pp. 182--191.

\bibitem{McMi}
B.~McMillan, ``Two inequalities implied by unique decipherability,'' \emph{IRE
  Trans. Inf. Theory}, vol. IT-2, no.~4, pp. 115--116, Dec. 1956.

\bibitem{MoTu97}
A.~Moffat and A.~Turpin, ``On the implementation of minimum redundancy prefix
  codes,'' \emph{IEEE Trans. Commun.}, vol.~45, no.~10, pp. 1200--1207, Oct.
  1997.

\bibitem{MTK}
A.~Moffat, A.~Turpin, and J.~Katajainen, ``Space-efficient construction of
  optimal prefix codes,'' in \emph{Proc., IEEE Data Compression Conf.}, Mar.
  28--30, 1995, pp. 192--202.

\bibitem{Newt}
I.~Newton, \emph{Opticks}.\hskip 1em plus 0.5em minus 0.4em\relax London, UK:
  Smith and Walford, 1704, available from
  \url{http://burndy.mit.edu/Collections/Babson/Online/Opticks}.

\bibitem{Reny}
A.~R{\'{e}}nyi, \emph{A Diary on Information Theory}.\hskip 1em plus 0.5em
  minus 0.4em\relax New York, NY: John Wiley {\&} Sons Inc., 1987, original
  publication: {\it Napl\`{o} az inform\'{a}ci\'{o}elm\'{e}letr\H{o}l},
  Gondolat, Budapest, Hungary, 1976.

\bibitem{Schi}
B.~Schieber, ``Computing a minimum-weight $k$-link path in graphs with the
  concave {Monge} property,'' \emph{Journal of Algorithms}, vol.~29, no.~2, pp.
  204--222, Nov. 1998.

\bibitem{Schw}
E.~S. Schwartz, ``An optimum encoding with minimum longest code and total
  number of digits,'' \emph{Inf. Contr.}, vol.~7, no.~1, pp. 37--44, Mar. 1964.

\bibitem{Tome}
I.~Tomescu, ``Optimum {Huffman} forests,'' \emph{J. Universal Comput. Sci.},
  vol.~3, no.~7, pp. 813--820, July 1997.

\bibitem{TuMo}
A.~Turpin and A.~Moffat, ``Practical length-limited coding for large
  alphabets,'' \emph{The Comput. J.}, vol.~38, no.~5, pp. 339--347, 1995.

\bibitem{TuMo2}
------, ``Efficient implementation of the package-merge paradigm for generating
  length-limited codes,'' in \emph{Proc., Computing: The Australasian Theory
  Symposium}, Jan. 29--30, 1996, pp. 187--195.

\bibitem{Leeu}
J.~{van Leeuwen}, ``On the construction of {Huffman} trees,'' in \emph{Proc.
  3rd Int. Colloquium on Automata, Languages, and Programming}, July 1976, pp.
  382--410.

\bibitem{WMB}
I.~H. Witten, A.~Moffat, and T.~Bell, \emph{Managing Gigabytes}, 2nd~ed.\hskip
  1em plus 0.5em minus 0.4em\relax San Francisco, CA: Morgan Kaufmann
  Publishers, 1999.

\end{thebibliography}

\ifx \cyr \undefined \let \cyr = \relax \fi

\end{document}